\documentclass[doublecol]{epl2} 

\usepackage{amsfonts, amsmath}
\usepackage{color, graphicx}

\def\gtrsim{\mathrel{\raise.4ex\hbox{$>$}\kern-0.8em\lower.7ex\hbox{$\sim$}}}
\def\lesssim{\mathrel{\raise.4ex\hbox{$<$}\kern-0.8em\lower.7ex\hbox{$\sim$}}}

\newcommand{\bk}{{\boldsymbol k}}

\newcommand{\be}{{\bf e}}

\newcommand{\da}{\downarrow}
\newcommand{\ua}{\uparrow}

\title{Emergent Spin Liquids in the Hubbard Model on the Anisotropic Honeycomb Lattice}

\author{Guangquan Wang\inst{1,2,3} \and Mark O. Goerbig\inst{1} \and Christian Miniatura\inst{2,3,4} \and Beno\^it Gr\'emaud\inst{2,3,5}}
\shortauthor{G. Wang \etal}

\institute{
\inst{1} Laboratoire de Physique des Solides, %
Univ.~Paris-Sud, CNRS UMR 8502, F-91405 Orsay, France\\
\inst{2} Centre for Quantum Technologies, %
National University of Singapore, Singapore 117543, Singapore\\
\inst{3} Department of Physics, Faculty of Science, %
National University of Singapore, Singapore 117542, Singapore\\
\inst{4} Institut Non Lin\'eaire de Nice, UNS,
CNRS; 1361 route des Lucioles, 06560 Valbonne, France\\
\inst{5} Laboratoire Kastler-Brossel, %
UPMC-Paris 6, ENS, CNRS; 4 Place Jussieu, F-75005 Paris, France
}

\pacs{71.30.+h}{Metal-insulator transitions and other electronic transitions}
\pacs{71.10.Fd}{Lattice fermion models}
\pacs{37.10.Jk}{Atoms in optical lattices}

\abstract{
We study the repulsive Hubbard model on an anisotropic honeycomb lattice within a mean-field and a slave-rotor treatment. In addition
to the known semi-metallic and band-insulating phases, obtained for very weak interactions, and the anti-ferromagnetic phase at 
large couplings, various insulating spin-liquid phases develop at intermediate couplings. Whereas some of these spin liquids have
gapless spinon excitations, a gapped one occupies a large region of the phase diagram and 
becomes the predominant phase for large hopping anisotropies. 
This phase can be understood in terms of weakly-coupled strongly dimerized states.
}

\begin{document}

\maketitle

One of the most salient features of graphene, an atomically thin graphite sheet with carbon atoms arranged in a 
honeycomb lattice, is certainly the relativistic character 
of its low-energy electronic excitations \cite{antonioRev}. An important issue of present research efforts is a deeper 
understanding of the role of interactions in this novel two-dimensional material and of the persistance of the 
semi-metallic (SM) phase when such interactions become relevant. As a consequence of the poor screening properties of 
electrons in the SM phase, the Coulomb interaction remains long-range and may be described in terms of an effective fine-structure 
constant that turns out to be density-independent, $\alpha_{\text{G}}=e^2/\hbar\epsilon v_{\text{F}}\simeq 2.2/\epsilon$
(for recent reviews on electronic interactions in graphene, see Refs. \cite{goerbigRev} and \cite{kotovRev}). 
Here $v_{\text{F}}\simeq c/300$
is the Fermi velocity in terms of the speed of light $c$, and $\epsilon$ is the dielectric function of the
surrounding medium. Whereas lattice-gauge theories predict a flow to strong coupling above some critical value 
$\alpha^c_{\text{G}}\sim 1.1$ \cite{drut}, 
even suspended graphene seems to be weakly correlated, with a stable SM phase \cite{Du}.
A reason for this effective flow to weak coupling
may be an intrinsic dielectric constant in graphene due to virtual interband excitations \cite{GGV}. Indeed, recent renormalization-group
studies confirm this picture of weakly-interacting electrons in graphene \cite{vlada}.

A perhaps more promising system for the study of the interplay between strong (short-range) correlations and the relativistic
character of Dirac fermions may well be a gas of cold fermionic atoms trapped in an optical honeycomb potential \cite{zhu,KL}. As compared to graphene, such a system has several advantages. First, neutral atoms
residing on the same site exhibit short-range interactions which can often be tuned over orders of magnitude and turned repulsive or attractive by using Feshbach resonances \cite{Bourdel}. Indeed, recent experiments have proven the feasibility of implementing
optical honeycomb lattices and probing interaction physics in the context of bosonic $^{87}$Rb atoms \cite{sengstock}.
Second, the hopping rates between neighboring sites are easily controlled by the laser configuration and beam intensities as both determine the depth and position of the optical potential wells. 
Rather moderate changes in the laser intensities can significantly imbalance the tunneling rates, 
leading to situations ranging from weakly-coupled zig-zag linear chains to weakly-coupled dimers \cite{KL}. This situation needs to be contrasted to graphene, where unphysically
large lattice distortions are required to obtain the limit $t'\sim 2t$ \cite{DefGraph}, where novel physical phenomena are expected
\cite{dietl}.

Starting from the repulsive fermionic Hubbard model (RFHM) for spin-1/2 particles
with onsite interactions $U\!\!>\!\! 0$ and identical nearest-neighbor hopping rates $t$, mean-field calculations for a 
half-filled lattice and zero temperature predict

a SM phase at small values of $U/t$ and an anti-ferromagnetic (AF) phase
developing above $U_{\text{c}}\simeq 2.2 t$ \cite{Sorella,Peres}. 
Quantum Monte Carlo (QMC) calculations confirm this SM-AF transition but at a higher transition point $U_\text c\sim 4.5 t$ \cite{Sorella,Meng} while dynamical-mean-field estimates yield even larger values, 
$U_\text c\sim 10 t$ \cite{Jafari}. Most saliently, recent QMC investigations have revealed an intriguing insulating spin-liquid (SL) phase, 
with localized charges but no spin ordering,
that emerges below the AF transition \cite{Meng} and that may be related to the exotic Mott insulator identified in 
slave-rotor studies \cite{Lee,hermele,Karyn}.
The transition points derived from the slave-rotor
theory \cite{Florens} also occur at globally smaller values of $U/t$ than in QMC calculations. 

\begin{figure}[t]
\includegraphics[width=0.45\textwidth]{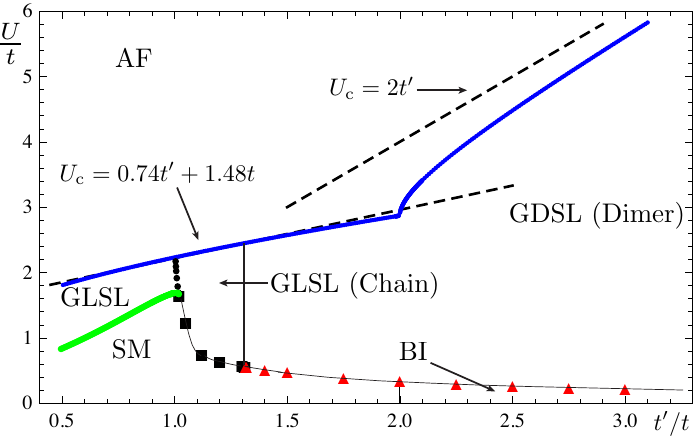}
\caption{(Color online) 
Mean-field phase diagram obtained for the anisotropic repulsive Hubbard model on the honeycomb lattice at zero temperature and half-filling. The upper thinner blue line delineates the AF phase, whereas the two linear functions (the dashed lines) for $U_\text c$ are obtained from simple scaling arguments (see text). The lower thick green line indicates a second-order phase transition between the SM and the gapless SL of the third solution (GLSL), whereas the dotted line are first-order transitions, which consists of three sections: the black round dots for $1.005\lesssim t'/t\lesssim1.02$ between the GLSL of the third solution and that of the chain solution, the black squares for $1.02\lesssim t'/t\lesssim1.31$ between the SM and the GLSL of the chain solution and, above $t'/t\simeq 1.31$, the red triangles between the SM (or BI) and a gapped SL (GDSL) that consists of decoupled dimers on the bonds with $t'$. The curve across the black squares and red triangles is drawn to direct the eyes.
}\label{fig:phasediag}%
\end{figure}

Here, we study the RFHM on a half-filled honeycomb lattice where one of the nearest-neighbor hopping parameters $t'$ is larger than the other two $t$. Such a 
lattice reveals an astonishingly rich phase diagram (see Fig.~\ref{fig:phasediag}), in which  
SL phases predominate at large values of $t'/t$.
At $U=0$, this system develops a topological phase transition between a semi-metal and a band insulator (BI). Indeed, at $t'=2t$, the two Dirac points responsible for the SM phase merge and eventually disappear while a band gap opens \cite{zhu,montambaux}. 
These topological properties are prominently reflected in the SL phase,
where the spin excitations acquire a gap as a function of renormalized hopping parameters. Finally, we show that the SL phase finds a
compelling interpretation in terms of weakly-coupled dimer states in the limit $t'\gg t$.

\begin{figure}[t]
\includegraphics[width=0.3\textwidth]{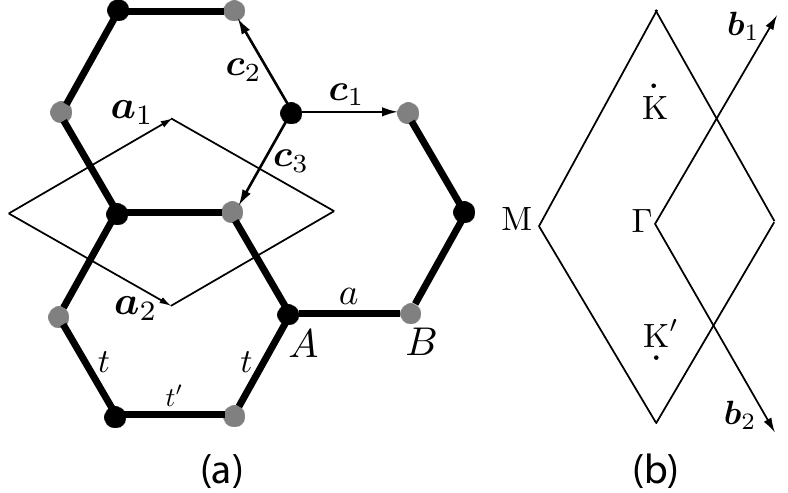}
\caption{(a): The honeycomb lattice and its diamond-shaped unit cell delineated by the Bravais lattice vectors $\boldsymbol{a}_{1,2}$. Black (resp. grey) sites belong to sublattice $A$ (resp. $B$), $a$ being the distance between nearest-neighboring sites. The hopping amplitude $t'$ connects sites linked by the vector ${\bf c}_{1}$ while $t$ connects sites linked by ${\bf c}_{2,3}$. (b): The reciprocal lattice is spanned  by the primitive vectors $\boldsymbol{b}_{1,2}$, and has the diamond-shaped primitive unit cell as shown. The $\Gamma$ point is the center of the unit cell. In the isotropic limit, the two inequivalent Dirac points are located at K and K$'$.
\label{fig:latticestru}}%
\end{figure}

As the honeycomb lattice is made of two shifted identical triangular sublattices $A$ and $B$, the kinetic term 
of the RFHM, $\mathcal{H}=\mathcal{H}_0+\mathcal{V}$, reads
\begin{equation}
\label{eq:modelH0}
\mathcal{H}_0=-t\sum_{\langle i,j\rangle\sigma}\hat{a}^{\dagger}_{i\sigma}\hat{b}_{j\sigma}-t'\sum_{\langle i',j'\rangle\sigma}\hat{a}^{\dagger}_{i'\sigma}\hat{b}_{j'\sigma} +(\mathrm{h.c.}),
\end{equation}
where $\hat a_{i\sigma}^{(\dagger)}$ and $\hat b_{j\sigma}^{(\dagger)}$ annihilate (create) a fermion with spin $\sigma=\ua,\da$ on the sites of the $A$ and $B$ sublattices, respectively. It also distinguishes the hopping amplitude $t'>0$, chosen along neighboring sites linked by the vector
${\boldsymbol c}_{1} = a \be_x$, which differs from the other two $t>0$ chosen along neighboring sites linked by ${\boldsymbol c}_{2}=a(-\be_x/2 + \sqrt{3}\be_y/2)$ or by ${\boldsymbol c}_{3}=-a(\be_x/2 + \sqrt{3}\be_y/2)$,
$a$ being the lattice constant [see Fig.~\ref{fig:latticestru}(a)]. At half-filling, the onsite interaction term reads
\begin{equation}
\label{eq:modelV}
\mathcal{V}=-\frac{U}{2}\sum_{i}(\hat{n}_{i\uparrow}-\hat{n}_{i\downarrow})^2,
\end{equation}
with $U>0$, the summation running over both sublattices 
and $\hat{n}_{i\sigma}=\hat a_{i\sigma}^{\dagger}\hat a_{i\sigma}$ or 
$\hat b_{i\sigma}^{\dagger}\hat b_{i\sigma}$ being the corresponding number operators. 
Here we consider a balanced population between spin-$\ua$ and spin-$\da$. When $U=0$, the Hamiltonian is
readily diagonalized in reciprocal space, and one obtains the two energy bands 
$\epsilon_{\pm}^{\boldsymbol{k}}=\pm|\gamma^{\boldsymbol{k}}|$, in terms of the weighted sum of phase factors
$\gamma^{\boldsymbol k} = t g^{\boldsymbol{k}} + t' g'^{\boldsymbol{k}}$ with $g^{\boldsymbol k} = e^{i\boldsymbol k\cdot\boldsymbol{c}_2}+e^{i\boldsymbol k\cdot\boldsymbol{c}_3}$ and $g'^{\boldsymbol k}=e^{i\boldsymbol k\cdot\boldsymbol{c}_1}$. When $t'=t$, one recovers the usual Dirac points
located at K and K$'$, as depicted in Fig.~\ref{fig:latticestru}(b). The Dirac points move towards the point M as $t'$ is increased, where they finally merge when $t'=2t$ \cite{zhu,montambaux}. 
At half-filling, this topological phase transition separates a SM phase (for $t'<2t$), with two Dirac points, from a BI (for $t'>2t$), the insulating gap $\Delta_{\text{I}}=t'-2t$ opening at the point M. 

For this half-filled lattice, one expects a Mott insulating (MI) phase at large $U$ in which each lattice site is exactly occupied by one atom. As a consequence of the superexchange interaction $4t^2/U$, the spins of the particles are further frozen in an AF N\'eel state that breaks the lattice inversion symmetry, with e.g.~spin-$\ua$ atoms on the $A$ sublattice and spin-$\da$ on the 
$B$ sublattice. This state is  described by the staggered mean-field order parameter,
$\langle \hat{n}_{i\uparrow}-\hat{n}_{i\downarrow}\rangle\equiv\Delta\ \text{if}\ i\in A,\ \text{and}\ -\Delta\ \text{if}\ i\in B$. The mean-field approximation $\hat{n}_{i\uparrow}\hat{n}_{i\downarrow}\approx [(1\pm\Delta)\hat{n}_{i\uparrow}+(1\mp\Delta)\hat{n}_{i\downarrow}]/2$ in the interaction term $\mathcal{V}$ leads to a quadratic Hamiltonian which is diagonalized through a Bogoliubov-Valatin transformation.
The corresponding excitation spectrum is $E^{\boldsymbol{k}}_{\pm}=\pm\sqrt{\Delta_{\text M}^{2}+|\gamma^{\boldsymbol{k}}|^{2}}$. The Mott gap is found to be $\Delta_\text M=U\Delta/2$ and is therefore intimitely
related to the AF order parameter, i.e.~it opens as soon as the AF order sets in. In the thermodynamic limit, the self-consistent gap equation at zero temperature then reads
\begin{equation}
\label{eq:gapeq}
\int_{\text{FBZ}} \frac {d^2{\bf k}} \Omega \frac{U}{\sqrt{U^{2}\Delta^{2}+4|\gamma^{\boldsymbol{k}}|^{2}}}= 1,
\end{equation}
where $\Omega=8\pi^2/(3\sqrt{3}a^2)$ is the area of the first Brillouin zone (FBZ).
The transition line $U_\text c$ delineating the AF phase (blue solid line in Fig.~\ref{fig:phasediag}) is obtained from Eq.~(\ref{eq:gapeq})
by setting $\Delta=0$. Notice that the value $U_\text{c}\simeq 2.2 t$, obtained at $t'=t$, agrees with former mean-field calculations \cite{Sorella,Peres}. The critical value $U_\text c$ is however shifted to larger values when $t'$  increases. This
shift may be understood from a simple scaling argument when considering the full band width $W= 2(2t+t')$ of the non-interacting case. Since all
particles are localized in the AF phase, one needs to consider states with energies up to $\sim W$. The critical value 
for the isotropic case $t'=t$ may now be expressed as $U_\text c/W \simeq 0.37$. If we consider this value to remain constant when varying
$t'$, one obtains $U_\text c\simeq 0.74t'+1.48 t$, a relation that describes to great accuracy the transition line for $t'\leq 2t$ (Fig.~\ref{fig:phasediag}). At larger 
values of $t'$, the mean-field Mott gap $\Delta_\text M$ competes with the band insulator gap $\Delta_\text I$, and the transition line may be
understood qualitatively as the line where both gaps are equal. This yields the asymptotic behavior
$U_\text c = 2(t'-2t)\sim 2t'$ in the large $t'$-limit (dashed line in Fig.~\ref{fig:phasediag}).

In order to decouple the Mott transition from the AF phase and to study possible intermediate phases, 
more sophisticated methods than a simple mean-field treatment are required. Apart from QMC calculations \cite{Meng},
an intermediate MI spin-liquid phase has been identified in the honeycomb lattice within a slave-rotor treatment
\cite{Lee,hermele,Karyn}, where the fermion operators $\hat{a}_{j\sigma}$ and $\hat{b}_{j\sigma}$ are viewed as products of
two auxiliary degrees of freedom, $\exp(i\theta_j)\hat{f}_{j\sigma}$. 
We adopt, here, a U(1) slave-rotor treatment
that is expected to provide qualitatively correct results within the mean-field approximation. However, if one aims at an effective
low-energy theory for the spinon degrees of freedom, one needs to take into account the coupling to an SU(2) gauge field as discussed 
in Ref.~\cite{hermele}.
The bosonic ``rotor'' field $\theta_j$ is conjugate to the
total charge at site $j$, described by the angular momentum $\hat{L}_j=i\partial_{\theta_j}$, and the fermion operator $\hat{f}_{j\sigma}$
carries the spin (``spinon''). As this procedure artificially enlarges the Hilbert space, double counting needs to be cured by imposing the constraint 
\begin{equation}
\hat L_j+\sum_{\sigma}\hat{f}^{\dagger}_{j\sigma}\hat{f}_{j\sigma}=\boldsymbol 1
\label{eq:constrantrho}
\end{equation} 
at each site $j$. Whereas the rotor and spinon fields are
coupled via the kinetic Hamiltonian $\mathcal{H}_0$, the interaction term is described
solely in terms of the rotor degrees of freedom, $\mathcal{V}=(U/2)\sum_j\hat{L}_j^2$, the particle-hole-symmetric form of which also renders the chemical potential $\mu=0$. 
Following Ref.~\cite{Karyn}, the rotor and spinon degrees of freedom may now be decoupled within a mean-field treatment by defining the averages $Q_\theta^{(\prime)}=\sum_{\sigma}\langle\hat{f}^{\dagger}_{i\sigma}\hat{f}_{j\sigma}\rangle$ and  $Q_f^{(\prime)}=\langle\exp(-i\theta_{ij})\rangle$, where $\theta_{ij}\equiv\theta_i - \theta_j$ and
$i$ and $j$ are nearest neighbors connected by $\boldsymbol{c}_{1}$ for the primed averages and by $\boldsymbol{c}_{2,3}$ otherwise. 
Notice that the mean-field parameters along $\boldsymbol{c}_{2,3}$ are assumed to be equal, hence the corresponding symmetries are not broken here.
The decoupled mean-field Hamiltonian may then be written as
$\mathcal{H}\simeq \mathcal{H}_\theta + \mathcal{H}_f$, with
\begin{subequations}
\begin{eqnarray}
\mathcal{H}_{\theta}&=&-t\sum_{\langle i,j\rangle}Q_\theta e^{-i\theta_{ij}}-t'\sum_{\langle i',j'\rangle}Q'_\theta e^{-i\theta_{i'j'}}+(\mathrm{h.c.})\nonumber\\
                          &&+\frac{U}{2}\sum_i\left(\hat L_i+\frac{h_i}{U}\right)^2,\\
\mathcal{H}_{f}&=&-t\sum_{\langle i,j\rangle\sigma}Q_f \hat{f}^{A\dagger}_{i\sigma}\hat{f}^B_{j\sigma}-t'\sum_{\langle i',j'\rangle\sigma}Q'_f \hat{f}^{A\dagger}_{i'\sigma}\hat{f}^B_{j'\sigma}+(\mathrm{h.c.})\nonumber\\
                          &&-\sum_{i\sigma}h_i\hat{f}^{\dagger}_{i\sigma}\hat{f}_{i\sigma},
\label{eq:fHamiltonian}
\end{eqnarray} 
\end{subequations}
where $h_i$ is a local Lagrange multiplier ensuring the constraint (\ref{eq:constrantrho}). At half-filling, particle-hole symmetry imposes that $h_i=0$. 
The Mott transition may now be interpreted in terms of rotor condensation. Indeed, in the rotor-condensed phase, the phase $\theta_j$ is fixed and the
number of particles (or the angular momentum $\hat{L}_j$) therefore fluctuates on the lattice sites. This corresponds 
to the SM phase for $t'\le2t$ or the BI for $t'>2t$. In the MI phase, however, there is no spin ordering since the spinon Hamiltonian $\mathcal{H}_f$ has no interaction term. 

\begin{figure}[t]
\includegraphics[width=0.4\textwidth]{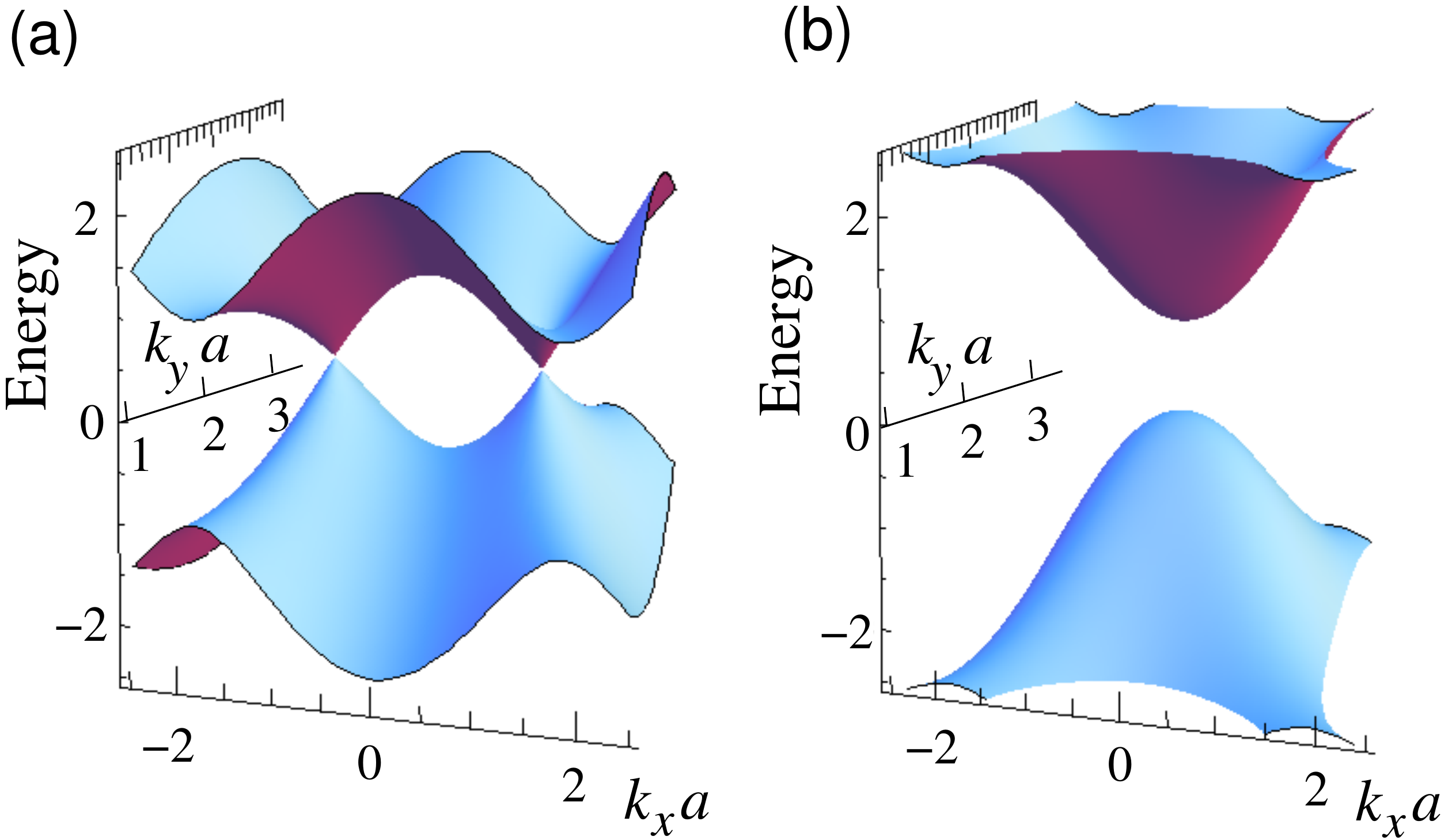}
\caption{(Color online) Spinon dispersion obtained from mean-field Hamiltonian $\mathcal{H}_f$. (a) For $t'Q_f'=tQ_f$, the dispersion consists of two distinct Dirac points, and one obtains a gapless
spin liquid.
(b) For $t'Q_f'=2.5 tQ_f$, the dispersion relation is gapped for any vector $\bk$ (gapped spin liquid).
\label{fig:spinon}}%
\end{figure}

Most saliently, $\mathcal{H}_0$ can be mapped onto $\mathcal{H}_f$ through $(t',t) \to (t'Q'_f,tQ_f)$, such that the low-energy spinon excitations are governed by the same topological properties as the non-interacting Dirac fermions. 
In particular, similarly to $\gamma^{\bk}$, we find the weighted sums of phase factors 
\begin{equation}\label{eq:phaseF}
\Gamma_{f}^{\boldsymbol{k}}=tQ_{f}g^{\boldsymbol{k}}+t'Q'_{f}g'^{\boldsymbol{k}}
\end{equation}
and
\begin{equation}\label{eq:phaseX}
\Gamma_{\theta}^{\boldsymbol{k}}=tQ_{\theta}g^{\boldsymbol{k}}+t'Q'_{\theta}g'^{\boldsymbol{k}}.
\end{equation}
Therefore, there exists a critical value 
\begin{equation}
t'Q_f' = 2tQ_f
\end{equation}
that separates a SL with gapless spinon excitations, around two distinct Dirac points at the Fermi level
(for $t'Q'_f < 2 tQ_f$), from a gapped SL phase (for $t'Q'_f >2 tQ_f$). The spinon dispersion calculated from the Hamiltonian 
$\mathcal{H}_f$ is depicted in Fig. \ref{fig:spinon} for two characteristic sets of parameters.

To obtain the Mott transition line, one needs to self-consistently solve for $Q_\theta$, $Q'_{\theta}$, $Q_f$, and $Q'_f$. This is achieved
with
the help of the imaginary-time action 
\begin{equation}
\label{Action}
S = \int_{0}^{\beta}d\tau\Big[\sum_{j}\Big(-iL_j\partial_{\tau}\theta_j+
\sum_{\sigma}f^{*}_{j\sigma}\partial_{\tau}f_{j\sigma}\Big)+\mathcal{H}\Big],
\end{equation}
where $\beta=1/(k_BT)$ is the inverse temperature and $k_B$ the Boltzmann constant. We next rewrite  \eqref{Action}  in terms of the fields $X_j = \exp(i\theta_j)$ and constrain the normalization $|X_j(\tau)|^2=1$ via a Lagrange multiplier $\rho$. The resulting Green's functions for the fields $X$ and $f$ are 
\begin{subequations}
\begin{eqnarray}
\tilde G^{-1}_\theta(\boldsymbol k, i\nu_n) &=& -(\nu_n^2/U+\rho-|\Gamma_\theta^{\boldsymbol k}|), \label{eq:bgreens}\\
\tilde G^{-1}_f(\boldsymbol k, i\omega_n) &=& i\omega_n-|\Gamma_f^{\boldsymbol k}|, 
\end{eqnarray}
\end{subequations}
where $\omega_n$ and $\nu_n$ are the fermionic and bosonic Matsubara frequencies, respectively. 
A change of $U\to U/2$ is performed in equation (\ref{eq:bgreens}), in order to preserve the correct atomic limit \cite{Florens}.
Based on the form of the rotor Green's function, one may define the charge gap as
\begin{equation}
\Delta_\text{g}=2\sqrt{U(\rho-|\Gamma_\theta^{\boldsymbol{k}}|_\text{max})},
\end{equation} 
where $|\Gamma_\theta^{\boldsymbol{k}}|_\text{max}$ is the maximum of $|\Gamma_\theta^{\boldsymbol{k}}|$ over the FBZ.
The normalization of the $X$-field yields the equation \cite{Florens} 
\begin{equation}
\int_{\text{FBZ}} \frac{d^2{\boldsymbol k}}{\beta\Omega}\sum_n \tilde G_\theta(\boldsymbol k, i\nu_n) =-1.
\label{eq:1stselfconsis}
\end{equation}

We consider the rotor-disordered and rotor-condensed phases separately.
In the former case, we perform the sum over Matsubara frequencies in Eq. (\ref{eq:1stselfconsis}) and consider the zero temperature limit. We find 
\begin{equation}
\int_{\text{FBZ}} \frac{d^2{\boldsymbol k}}{\Omega}\frac{U}{\sqrt{4U(\rho-|\Gamma_\theta^{\boldsymbol{k}}|)}}=1,
\label{eq:eqforrho}
\end{equation}
which, together with the equations for the $Q$s, completes the set of self-consistency equations for the $Q$s and $\rho$ for the rotor-disordered (MI) phase.
For the rotor-condensed (SM/BI) phase, as in the case of a normal Bose-Einstein condensate, 
a macroscopic fraction, namely $n_0$  per lattice site, of the particles occupies the $\boldsymbol{k}=0$ state, 
which corresponds to the lowest energy $-|\Gamma_\theta^{\boldsymbol{k}}|_\text{max}$.
The chemical potential ($-\rho$) is equal to this lowest energy.
The corresponding equation becomes
\begin{equation}
2n_0+\int_{\text{FBZ}} \frac{d^2{\boldsymbol k}}{\Omega}\frac{U}{\sqrt{4U(|\Gamma_\theta^{\boldsymbol{k}}|_\text{max}-|\Gamma_\theta^{\boldsymbol{k}}|)}}=1,
\label{eq:eqforn0}
\end{equation}
which, instead of Eq. (\ref{eq:eqforrho}), completes the set of self-consistent equations for the $Q$s and $n_0$ for the 
rotor-condensed (SM/BI) phase.

In the rotor-disordered (MI) phase, one has $\Delta_\text{g}>0$ ($\rho>|\Gamma_\theta^{\boldsymbol{k}}|_\text{max}$) and $n_0=0$, while in the rotor-condensed (SM/BI) phase, one has $n_0>0$ and $\Delta_\text{g}=0$. A second-order transition line is defined by $\Delta_\text{g}=0$ and $n_0=0$. As we show below, part of the phase transition curve that we obtained is of first order, indicated by a jump of $\Delta_\text{g}$ or $n_0$ from some finite values to zero. For the rotor-disordered (MI) phase, one finds three distinct solutions with different types of spinon excitations:

The \textit{chain solution} has vanishing mean-field parameters on the horizontal bonds, i.e.~$Q'_f=Q'_\theta=0$, and thus describes a system composed of decoupled vertical zig-zag chains [corresponding to the bonds with a hopping $t$ in Fig. 
\ref{fig:latticestru}(a)].
A self-consistent calculation yields $Q_\theta=-2/\pi\approx -0.637$, while $Q_f$ and $\rho$ are functions of $U$. 
The associated SL phase in this case is gapless, since $t'Q'_f=0<2tQ_f$.

The \textit{dimer solution} has $Q_f=Q_\theta=0$, and hence describes a system composed of decoupled dimers
on the horizontal bonds of the honeycomb lattice [see Fig. \ref{fig:latticestru}(a)].
Self-consistency requires that $Q'_f=0.5$ and $Q'_\theta=-1$. One may easily verify that $\Delta_\text{g}=U$ for this solution, independent of $t'$. This relation also fixes the value of $\rho$.
This SL is gapped in the spinon channel, since $t'Q'_f=0.5t'>2tQ_f=0$. 

A {\sl third (more general) solution} in the rotor-disordered SL phase is continuously connected to the rotor-condensed SM or BI 
phase, and it therefore separated from the latter by a second-order phase transition. This solution may be obtained by 
starting in the rotor-condensed phase with a sufficiently small value of $U$, where the mean-field parameters are obtained
numerically. These parameters serve as the starting point for the next calculation step (i.e.~for $U+\epsilon$, with 
an incremental energy step $\epsilon>0$). One thus obtains a continuous line of solution in parameter space. 
The condensate fraction $n_0$ decreases with increasing on-site
repulsion $U$, and the rotor-disordered MI phase is obtained when $n_0<0$. Furthermore, the charge gap $\Delta_\text{g}$ increases from zero continuously when $U$ is increased, as one expects for a second-order phase transition.

In order to identify which of the above-mentioned solutions is chosen by the system for a particular set of parameters ($t'/t,U/t$),
we have performed a free-energy analysis of the three solutions. The free energy per lattice site reads $\mathcal{E}=-(k_BT/2N_\text{c})\ln\mathcal{Z}$
in terms of the partition function $\mathcal{Z}$ and the number of lattice sites $2N_\text c$, and one has, 
in the zero-$T$ limit, in which the free energy becomes the internal energy
\begin{subequations}
\begin{equation}
\mathcal{E}=\mathcal{E}_f+\mathcal{E}_\theta-[2tQ_f Q_\theta+t'Q'_f Q'_\theta+\rho],
\label{eq:FE}
\end{equation}
where
\begin{equation}
\mathcal{E}_f=-\int_{\text{FBZ}}\frac{d^2{\boldsymbol k}}{\Omega}|\Gamma_f^{\boldsymbol{k}}|
\end{equation}
is the free energy of the fermionic part, and
\begin{equation}
\mathcal{E}_\theta=\int_{\text{FBZ}}\frac{d^2{\boldsymbol k}}{2\Omega}\sqrt{U(\rho-|\Gamma_\theta^{\boldsymbol{k}}|)}
\end{equation}
\end{subequations}
that of the bosonic part.
Note that in the condensed phase, $\rho=|\Gamma_\theta^{\boldsymbol{k}}|_\text{max}$.
The last term in equation (\ref{eq:FE}) accounts for the correction to the free energy due to the dynamically unimportant constants, which have been ignored so far. 

\begin{figure}[t]
\includegraphics[width=0.4\textwidth]{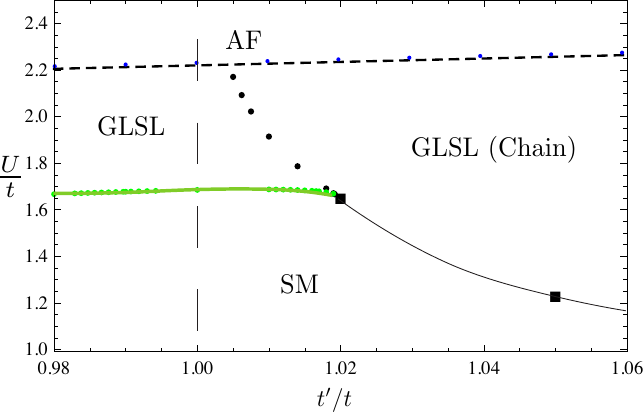}
\caption{(Color online) Zoom on the phase diagram in the vicinity of $t'=t$. For the notation and meaning of the symbols, refer to Fig.~\ref{fig:phasediag}. The dashed vertical line denotes the isotropic limit $t'=t$.
\label{fig:Isotropicphasediag}}%
\end{figure}

For a particular pair of $(t'/t,U/t)$, the true ground state is the one with the smallest $\mathcal{E}$.
The transition curves resulting from this free energy analysis are plotted in Fig.~\ref{fig:phasediag} (lower thicker green and dotted line), and Fig.~\ref{fig:Isotropicphasediag} for the vicinity of the isotropic limit ($t'=t$). One finds that, for $t'\lesssim1.005t$, the system undergoes a second-order phase transition (within the third solution) from the SM phase to a gapless SL phase (GLSL in Fig.~\ref{fig:phasediag}) upon increase of $U/t$;
for $1.005t\lesssim t'\lesssim1.019t$, 
the system first experiences the above-mentioned second-order phase transition from a SM to the GLSL (third solution), and then a first-order phase transition to another gapless one of the chain solution [GLSL (Chain)]; for $1.019t\lesssim t'\lesssim1.31t$, the 
second-order phase transition disappears and the system jumps directly from a SM phase to the GLSL (Chain)
via a first order phase transition. Finally, for $t'\gtrsim1.31t$, the system undergoes a first-order phase transition from the SM/BI phase to a gapped spin liquid of the dimer solution [GDSL (Dimer)]. The AF transition, which cuts the GLSL-GLSL (Chain) transition at $t'\simeq 1.005t$, but which is otherwise well above the SM/BI-SL transitions described here, is not treated within the current slave-rotor description. In principle, this AF transition may be described within a slave-rotor theory by reintroducing spin-charge correlations into the Hamiltonians (\ref{eq:fHamiltonian}) \cite{Zhao}.

For the isotropic case $t'=t$, our result of the second-order transition point $U_\text c/t\simeq 1.68$ agrees with Refs.~\cite{Lee,Karyn}. 
In this limit, the free-energy analysis indicates that the gapless SL of the third solution is the lowest while the gapped SL of the dimer solution is the highest in energy among the three solutions. 
This is in contrast to the QMC calculations \cite{Meng}, where a gapped SL was identified as the true ground state.
However, only an upper bound of the free energy of the dimer solution is provided by our free energy analysis.
This is especially the case in the isotropic limit, where a discrimination among the three hopping parameters is unphysical.
Indeed, possible kinetic dimer terms (e.g. resonating dimer moves around a hexagon) could lower the energy of the gapped SL of the dimer solution \cite{MSC01}.  
At a general value of $t'$, free energies are modified by these terms such that they are adiabatically connected to the (also modified) isotropic case. Similarly, the gapless chain solution may be unstable to an infinitesimal interchain coupling that, in the
context of the square lattice, is known to open a spin gap \cite{unstablechains}. An analysis of kinetic dimer terms and the stability
of the chain solution is, however, beyond the scope of this paper.

The predominating SL phase in the limit $t'\gg t$ is described by the dimer solution, where both $Q_f$ and $Q_X$ are zero. This in effect is equivalent to setting $t=0$, such that only sites connected by horizontal bonds are coupled by tunneling (Fig.~\ref{fig:latticestru}) and the whole system reduces to a set of decoupled dimers. It is therefore sufficient
to solve the Hubbard Hamiltonian for two interacting fermions occupying two lattice sites
coupled by the hopping amplitude $t'$. The two-particle Hilbert space obviously reduces to six states as the Pauli principle forbids fermions with identical spin to occupy a same site. By the same token, states where fermions with identical spin occupy different sites are obviously zero-energy eigenstates as hopping is Pauli-blocked. The relevant two-particle Hilbert space is thus spanned by
$|\ua,\da\rangle$, $|\ua\da,0\rangle$,
$|0,\ua\da\rangle$, and $|\da,\ua\rangle$, where the first entry describes the occupancy of the $A$ site and the second one that of the $B$ site.
The diagonalization of the Hubbard Hamitonian
then yields the ground state
\begin{equation}
|\psi_{\text{gs}}\rangle = \sin\phi \ |\mathcal{S}\rangle 
+\cos\phi \ \frac{|\ua\da,0\rangle + |0,\ua\da\rangle}{\sqrt{2}},
\end{equation}
where $|\mathcal{S}\rangle\equiv(|\ua,\da\rangle -|\da,\ua\rangle)/\sqrt{2}$ is the dimer singlet and $\phi= (2\alpha+\pi)/4$ with $\tan\alpha=U/4t'$.
The first excited state is the triplet state $|\mathcal{T}\rangle = (|\ua,\da\rangle + |\da,\ua\rangle)/\sqrt{2}$ and is separated from $|\psi_{\text{gs}}\rangle$ by the gap $\Delta_{\text{ST}}=[\sqrt{U^2+(4t^{\prime})^2}-U]/2$. The ground state is therefore
essentially the singlet state $|\mathcal{S}\rangle$, with an admixture of states with double occupancy the weight of which ($\cos\phi$)
vanishes in the 
large-$U/t'$ limit, where the gap is dominated by the exchange energy $\Delta_{\text{ST}}\approx  4t^{\prime 2}/U$. Because
$|\psi_{\text{gs}}\rangle$ remains the ground state over the whole range of values of $U/t'$, there is just one single phase
that continuously connects the BI to the GDSL phase. Notice that, within the mean-field U(1) slave-rotor treatment, 
there is not such a continuous connection because $\alpha$ can only take the values $0$ (for the BI) and $\pi/2$ (for the GDSL), 
double occupancy of a single site being ruled out in the MI phase. Furthermore, only at the 
asymptotic point, $U=\infty$, do the singlet and the triplet states become degenerate, and the AF state,  which requires a 
degeneracy (and a superposition) of $|\mathcal{S}\rangle$ and $|\mathcal{T}\rangle$,
may be formed. In the intermediate region,
the gap $\Delta_{\text{ST}}$ protects the ground state $|\psi_{\text{gs}}\rangle$ which is only marginally perturbed by a small inter-dimer coupling mediated
by $t\ll t'$. Within this simple dimer picture,
one may therefore understand the gapped SL [GDSL (Dimer)] in Fig.~\ref{fig:phasediag} as being adiabatically
connected to the state $|\psi_{\text{gs}}\rangle$.

In conclusion, we have investigated the repulsive fermionic Hubbard model on the anisotropic honeycomb lattice. This system could be experimentally realized by loading ultracold fermions at half-filling in a honeycomb optical lattice. Beside the SM, BI, and AF phases, which are 
readily obtained at the mean-field level, we have used a slave-rotor description to show the emergence of various SL
phases. Two gapless SL phases may be found from a self-consistent mean-field treatment of the slave-rotor theory: a chain solution that consists of essentially decoupled zig-zag chains and a phase that is connected continuously to the small-$U$ SM/BI phase. The latter
is found in an intermediate coupling regime $1.68\lesssim U/t\lesssim 2.2$ for the isotropic case, $t'=t$. Most saliently, a gapped dimer 
SL phase may be stabilized at rather low hopping anisotropies, $t'\gtrsim 1.31$, and is adiabatically connected to the large-$t'/t$ limit,
where the system may be described in terms of essentially decoupled dimer states. One may speculate that the dimer picture, if modified
by an additional kinetic term that allows for dimer moves, yields insight also into the physical properties of the system in the 
isotropic limit $t'=t$.

\acknowledgments

We thank A. H. Castro Neto, F. Cr\'epin, B.-G. Englert, J.-N. Fuchs, M. Gabay, N. Laflorencie, K. Le Hur, G. Montambaux, F. Pi\'echon, and M. Rozenberg for fruitful discussions. We acknowledge support from the France-Singapore Merlion program (CNOUS grant 200960 and FermiCold 2.01.09) and the CNRS PICS 4159 (France). The Centre for Quantum Technologies is a Research Centre of Excellence funded by the Ministry of Education and the National Research Foundation of Singapore.

\end{document}